\documentclass[10pt,conference,letterpaper]{IEEEtran}
\usepackage{cite}
\usepackage{amsmath,amssymb,amsfonts}
\usepackage{graphicx}
\usepackage{booktabs}
\usepackage{placeins}
\usepackage{xspace}
\graphicspath{{figures/}}
\newcommand{\sys}{HyCoInfer\xspace}
\newcommand{\HyCoInferAblationWidth}{.8}
\newcommand{\HyCoInferOverviewWidth}{.8}

\begin{document}

\title{HyCoInfer: Cloud--Edge Collaborative Inference for Large Language Models on Heterogeneous GPU--FPGA Architectures}

\author{\IEEEauthorblockN{Zewen Huang, Xiangwen Zhuge, Yingkun Hu, Zheng Yang, Qiang Ma}
\IEEEauthorblockA{Tsinghua University, China}}

\maketitle

\begin{abstract}
Large language model (LLM) serving relies predominantly on cloud GPUs, yet autoregressive decoding is memory-bound and often leaves their arithmetic units underutilized. Edge FPGAs offer attractive decoding efficiency, but are ill suited to sustaining a high-concurrency serving workload alone. This paper presents \sys, a heterogeneous cloud--edge serving system that assigns prefill to cloud GPUs and distributes decoding across cloud GPUs and edge FPGAs. The design addresses three barriers to practical offloading: capacity balancing across heterogeneous accelerators, the large KV cache crossing a constrained wide-area link, and the latency added by heterogeneous execution. \sys incorporates FPGA decode capacity into GPU prefill/decode role partitioning and places requests according to queue load, prompt length, and network state. It further combines layer-wise pipelined transfer, decoder-side precomputation, and bandwidth-adaptive KV cache compression based on residual vector quantization. We implement \sys using two NVIDIA V100 GPUs and an AMD Xilinx Alveo U280 and evaluate it over 100--1000 Mbps links. At 500 Mbps, \sys delivers up to 4.11$\times$ the request goodput of cloud-only vLLM and reduces mean end-to-end latency by up to 61.1\%. Across different network conditions and workloads, the system improves overall serving capacity, with the largest gains on communication-intensive tasks, and reduces per-request energy.
\end{abstract}


\section{Introduction}
As demand for interactive LLM services continues to grow, online inference systems face increasing workload pressure. Improving resource utilization, throughput, and latency under limited resource and cost budgets has therefore become a central problem in LLM serving.

Most cloud computing clusters use GPUs as their primary computing units, and existing inference systems accordingly focus on optimizing cloud GPU clusters. Despite their mature software ecosystem and high parallel throughput, GPUs are not utilized uniformly or fully across the two phases of LLM inference. Prompt processing (\emph{prefill}) computes many input tokens in parallel and can effectively use GPU arithmetic units. Autoregressive \emph{decoding}, however, generates one token at a time with low arithmetic intensity and repeatedly accesses model weights and a growing key-value (KV) cache. Its execution is therefore commonly constrained by memory bandwidth. Using GPUs for decoding leaves part of their compute capacity underutilized, limiting serving performance and the number of users that the system can support~\cite{zhongDistServeDisaggregatingPrefill2024,patelSplitwiseEfficientGenerative2024}.

Prior work has explored LLM inference on alternative hardware. FPGAs are attractive candidates for edge deployment because their customized datapaths and memory systems can improve execution efficiency. Existing FPGA LLM accelerators demonstrate competitive throughput and substantially better energy efficiency for small-batch inference; for example, FlightLLM achieves efficient decode with a TPOT of approximately 20 ms on an FPGA~\cite{zengFlightLLMEfficientLarge2024a}. However, the limited capacity of edge FPGAs makes them less capable than high-end GPUs on compute-intensive prefill and unsuitable for sustaining a high-concurrency serving workload alone.

The suitability of FPGAs for decoding motivates a hybrid design in which cloud GPUs process prefill, while cloud GPUs and edge FPGAs jointly undertake decoding. This design both releases cloud resources from part of the decoding workload and uses FPGAs in the phase where their efficiency advantage is most pronounced.

Realizing this design presents three systems challenges. First, the architecture requires coordinated inference across cloud nodes, edge nodes, heterogeneous hardware, and multiple network tiers. Second, serving workloads exhibit diverse request rates, prompt lengths, and available accelerator capacities, requiring efficient scheduling and request routing. Third, the limited bandwidth between cloud and edge can make KV-cache transfer a performance bottleneck and inflate time to first token (TTFT).

\begin{figure*}[t]
  \centering
  \includegraphics[width=\HyCoInferOverviewWidth\textwidth]{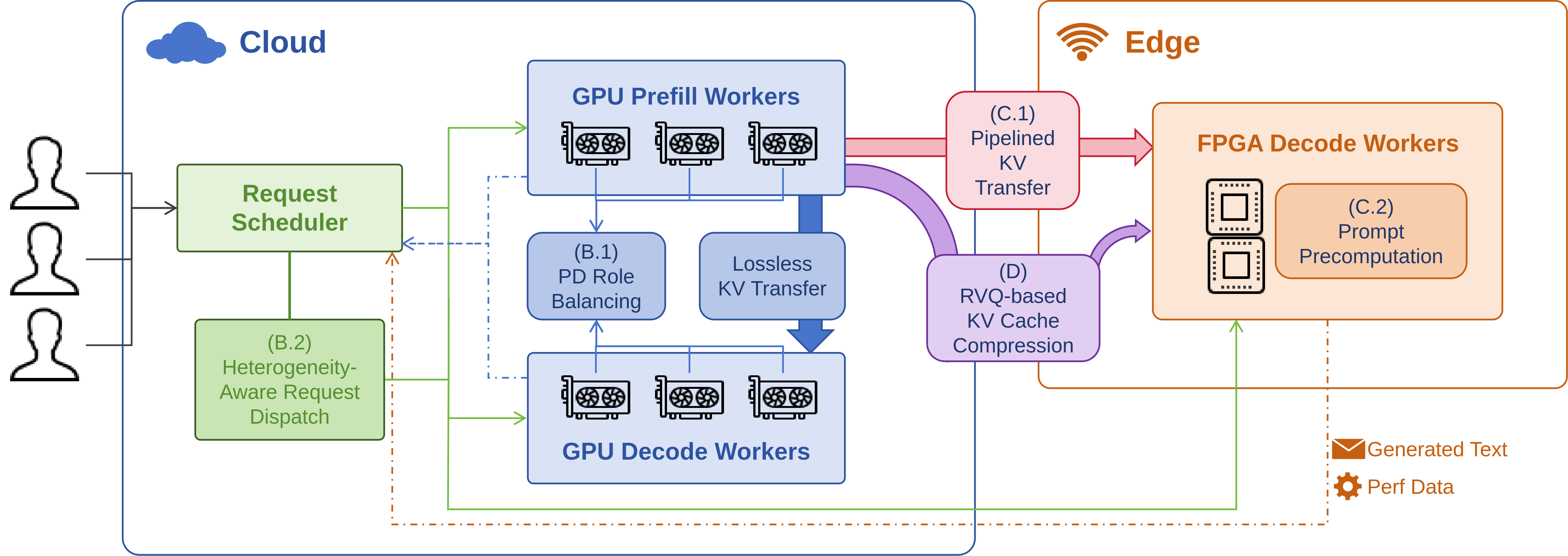}
  \caption{\sys architecture. Cloud GPUs perform prefill and share decoding with edge FPGAs; only compressed KV caches assigned to edge decoders cross the cloud--edge link.}
  \label{fig:overview}
\end{figure*}

To address these challenges, we present \sys, an adaptive GPU--FPGA collaborative inference system that improves serving capacity and responsiveness by jointly utilizing cloud and edge resources. As Fig.~\ref{fig:overview} shows, \sys builds on prefill--decode (PD) disaggregation. Using GPU and FPGA performance profiles, it accounts for the additional decode throughput supplied by FPGAs when partitioning GPUs between prefill and decode, thereby balancing the two stages. Requests are dispatched to GPU or FPGA decoders: short prompts favor the FPGA when its queue and network can satisfy the service-level objective (SLO), whereas long prompts preferentially remain in the cloud. For edge-bound requests, residual vector quantization (RVQ) converts KV vectors into compact codebook indices and adapts the number of residual stages to link bandwidth, enabling a progressive quality--latency tradeoff. Prefill-side layer-wise transmission overlaps communication with subsequent computation, while edge-prefix precomputation uses an otherwise idle FPGA compute window under low load.

We implement \sys on NVIDIA V100 GPUs and an AMD Xilinx Alveo U280 FPGA and evaluate it using four serving workloads, four quality benchmarks, and three link rates. At 500 Mbps on Qasper, \sys improves request goodput by 310.9\% and reduces latency by 56.2\% compared with cloud-only vLLM. Compared with unoptimized cloud--edge collaborative inference, it improves request goodput by 45.5\% and reduces latency by 14.4\%.

This paper makes three contributions. (1) We present \sys, a cloud--edge collaborative inference system for a hybrid GPU--FPGA architecture. Built on a mature high-performance inference framework, it provides strong performance and scalability. (2) We develop a scheduling strategy that combines FPGA-aware GPU PD balancing with prompt- and load-aware decoder selection, exploiting heterogeneous decode capacity while maintaining inter-stage throughput balance. Dynamic RVQ and layer-wise pipelining reduce cross-tier KV-transfer latency, while edge-prefix precomputation is designed to use idle FPGA capacity under low load. (3) On the three communication-intensive serving workloads at 100 Mbps, \sys improves request goodput by 66.8--233.9\% and reduces latency by 26.9--53.1\% over cloud-only deployment, while also delivering better energy efficiency.

\section{Preliminaries}
\subsection{Prefill, Decode, and PD Disaggregation}
LLM inference typically consists of two phases: prefill and decode. The prefill phase processes the complete input sequence, computes the hidden states of all input tokens across Transformer layers in parallel, and generates and stores the KV cache required by subsequent inference. This phase exposes substantial matrix-computation parallelism and is primarily compute-bound. The decode phase generates output tokens autoregressively. Each iteration reads the model weights and the accumulated KV cache to produce one new token. Because a decode iteration exposes limited computational parallelism, it is usually memory-intensive and more strongly constrained by memory bandwidth.

The distinct performance characteristics of prefill and decode lead to different resource requirements. When both workloads share an accelerator under continuous request arrivals, a long prefill can block active decode requests, while the sustained memory traffic of decoding can interfere with prefill. Such interference causes latency variance and degrades resource utilization.

Prefill--decode (PD) disaggregation deploys the two phases on separate workers and independently schedules and scales them according to their computational characteristics~\cite{zhongDistServeDisaggregatingPrefill2024,patelSplitwiseEfficientGenerative2024}. Once a prefill worker finishes processing the input, it transfers the request's KV cache to a decode worker, which continues generating output. This architecture reduces interference and permits hardware to be provisioned separately for compute-intensive prefill and memory-intensive decode. However, PD disaggregation also introduces KV-cache migration into the request critical path, especially when inter-worker bandwidth is constrained.

\subsection{Heterogeneous Hardware for LLM Inference}
GPUs integrate many parallel compute units and mature high-performance operator libraries, providing high throughput for large matrix multiplications and batched workloads. They are therefore well suited to compute-intensive prefill. Increasing batch size or processing many input tokens concurrently allows a GPU to utilize its arithmetic resources effectively. In contrast, autoregressive decode produces only one token per iteration, exposes limited parallelism, and repeatedly accesses model weights and the KV cache, making it difficult to exploit the GPU's peak compute capability.

FPGAs allow model operators to be mapped onto customized datapaths, on-chip buffers, and pipelines, and permit memory accesses to be optimized for low-precision computation. For small-batch, low-parallelism, and memory-intensive decode, an FPGA can reduce scheduling and data-movement overheads found in general-purpose accelerators and improve energy efficiency. FlightLLM reports that, when running Llama 2 7B on an AMD Xilinx Alveo U280, its TPOT is approximately 21.50 ms and reaches the A100 level~\cite{zengFlightLLMEfficientLarge2024a}. Resource constraints, however, make its TTFT nearly 30$\times$ that of the A100. This asymmetry further motivates using the FPGA for decode rather than standalone prefill.

\subsection{KV Cache}
During autoregressive Transformer inference, every layer caches the key and value tensors of preceding tokens. When a new token is generated, attention directly reads these cached tensors and avoids recomputing the intermediate states of all previous tokens. For a given model, each request's KV-cache size grows approximately linearly with its sequence length. As context windows and request concurrency increase, the aggregate memory footprint grows rapidly and limits the number of requests that can be served simultaneously~\cite{liuKIVITuningFreeAsymmetric2024}.

The KV cache is not only a local storage concern but also a communication burden in a PD-disaggregated system. Before decoding begins, the prefill worker must transfer the generated cache to the selected decode worker, and the transferred volume grows with input length. High-speed datacenter interconnects may hide part of this overhead; over a low-bandwidth or variable cloud--edge link, KV-cache transfer can instead dominate TTFT and offset the decoding benefit of heterogeneous hardware.

Existing work reduces KV-cache overhead through quantization, sparsification, and cache eviction. Quantization represents keys and values at lower numerical precision~\cite{liuKIVITuningFreeAsymmetric2024,hooperKVQuant2024}. Sparsification or token-selection techniques retain only states that are more important to subsequent attention, while eviction policies remove less valuable historical states according to attention scores, recency, or a capacity budget~\cite{zhangH2OHeavyHitterOracle2023,liSnapKVLLMKnows2024}. These approaches reduce storage or transferred bytes, but generally trade compression ratio against processing overhead and model quality.

\section{Method}
\subsection{System Model and Objective}
\subsubsection{System Architecture}
\sys serves online LLM requests using a cloud GPU cluster and an edge FPGA cluster under prefill--decode (PD) disaggregation. It comprises GPU prefill workers $P$, GPU decode workers $D$, FPGA decode workers $F$, and a global scheduler. A prefill worker computes the complete prompt and produces the per-layer KV cache. A GPU decoder receives a lossless cache over a high-speed cloud link, whereas an FPGA decoder receives a compressed cache over a bandwidth-constrained wide-area link. The scheduler tracks worker roles, capacities, in-flight requests, and per-link state, and selects the prefill worker, decode worker, and KV-transfer parameters.

We denote a deployment by $xP$-$yD$-$zF$, where $x$, $y$, and $z$ are the numbers of GPU prefill, GPU decode, and FPGA decode instances. Each request is prefilled exactly once and its KV cache is sent only to the selected decoder rather than broadcast. The GPU path uses lossless passthrough, while the FPGA path applies residual vector quantization (RVQ). Target-specific codecs avoid imposing RVQ overhead or accuracy loss on cloud GPU requests. \sys is built on vLLM V1. Every prefill producer connects to all GPU and FPGA endpoints, while request-level parameters select one destination and codec without changing the data-plane protocol when $x$, $y$, or $z$ changes.

\subsubsection{Request and Performance Model}
A request $r$ is represented by prompt length $n_r$, maximum output length $o_r$, and arrival time $a_r$. Its TTFT is decomposed as
\begin{equation}
T_r^{\mathrm{TTFT}}=T_r^{q,P}+\max(T_r^{P\text{-}\mathrm{KV}},T_r^{\mathrm{edge}})
+T_r^{q,D}+T_r^{D,1},
\label{eq:ttft-model}
\end{equation}
where $T_r^{q,P}$ and $T_r^{q,D}$ are prefill and decode queueing delays, $T_r^P$ is GPU prefill time, $T_r^{\mathrm{KV}}$ is the duration of KV encoding, transfer, reconstruction, and insertion, and $T_r^{D,1}$ is the time to produce the first token. $T_r^{P\text{-}\mathrm{KV}}$ denotes the joint critical path from the start of prefill until the target-side KV cache is ready, with
\begin{equation}
\max(T_r^P,T_r^{\mathrm{KV}})\le T_r^{P\text{-}\mathrm{KV}}
\le T_r^P+T_r^{\mathrm{KV}}.
\label{eq:prefill-kv-bounds}
\end{equation}
If KV migration starts only after the complete prefill, this term equals the serial upper bound. Layer-wise transfer instead overlaps the KV path with subsequent prefill layers and moves it toward the lower bound, while still retaining pipeline startup and tail costs. This overlap represents an additional optimization opportunity.

$T_r^{\mathrm{edge}}$ is the edge-prefix precomputation time introduced by \sys. The FPGA generates part of the prefix KV locally in parallel with the cloud prefill and suffix-KV path, so it contributes through a maximum with $T_r^{P\text{-}\mathrm{KV}}$ rather than a sum. Section~III-C2 presents this mechanism; when edge precomputation is disabled, $T_r^{\mathrm{edge}}=0$.

For worker $j$, the scheduler maintains in-flight count $q_j$, concurrency capacity $c_j$, and normalized load $\rho_j=q_j/c_j$. GPU capacity follows vLLM's maximum number of concurrent sequences; FPGA capacity follows the hardware pipeline. Each FPGA endpoint has an independent bandwidth $B_j$, rather than sharing a single abstract edge-bandwidth budget.

Under this model, \sys addresses a joint scheduling- and execution-level task. At the scheduling level, it first incorporates FPGA decode capacity into GPU PD partitioning to balance prefill and aggregate decode throughput (Sec.~III-B1), and then routes each request to a GPU or FPGA decoder based on request characteristics, worker load, and capacity (Sec.~III-B2). At the execution level, its optimizations center on the KV cache: decoder-side prefix precomputation and bandwidth-adaptive compression reduce cross-domain traffic volume (Secs.~III-C2 and III-D), while prefill-side layer-wise pipelining hides KV-transfer time (Sec.~III-C1).

\subsection{Scheduling-Level Optimizations}
\subsubsection{FPGA-Aware GPU PD Balancing}
Conventional PD partitioning considers only GPU performance in prefill and decode. To accommodate FPGAs, \sys treats GPUs as a homogeneous pool and explicitly accounts for FPGA decode capacity when assigning them to the two stages. Let the profiled throughputs of one GPU prefill, GPU decode, and FPGA decode instance be $r_P$, $r_D$, and $r_F$. A deployment has $N$ GPUs and $z$ FPGAs, with $x$ GPUs assigned to prefill and $y=N-x$ to decode. End-to-end throughput is bounded by the slower of prefill and aggregate decode. Avoiding persistent idleness or backlog therefore requires
\begin{equation}
xr_P\simeq yr_D+zr_F,\qquad x+y=N.
\label{eq:pd-balance}
\end{equation}
Solving gives
\begin{equation}
x^*=\frac{Nr_D+zr_F}{r_P+r_D},\qquad
y^*=\frac{Nr_P-zr_F}{r_P+r_D},
\label{eq:pd-ratio}
\end{equation}
after rounding and clipping to feasible instance counts. As $zr_F$ grows, fewer GPU decoders are required and more GPUs can be assigned to prefill.

A deployment can first profile the target model under representative sequence lengths and batching configurations and then use Eq.~\ref{eq:pd-ratio} to choose its $xP$-$yD$-$zF$ partition. A long-running service may also observe realized prefill and aggregate decode throughput over a coarse timescale and recalibrate the profiles or partition when necessary.

\subsubsection{Heterogeneity-Aware Request Dispatch}
Because edge KV traffic grows nearly linearly with prompt length, \sys jointly considers prompt size, normalized load, capacity, and recent SLO feedback. Before tokenization, online serving uses message character count as a low-cost estimate $\hat n_r$. Given an administrative threshold $n_0$ or observed mean $\bar n$, the effective threshold is $\theta=\alpha n_0$ or $\theta=\alpha\bar n$. After a request completes, $\alpha\leftarrow\min(1,\alpha+0.1)$ when both SLOs hold and $\alpha\leftarrow\max(0,\alpha-0.2)$ otherwise. The asymmetric update quickly reduces long edge placements after a violation.

The scheduler first finds the least-loaded GPU and FPGA endpoints $d_g^*$ and $d_f^*$. With no history, it conservatively chooses a GPU. If $\hat n_r\ge\theta$, it prefers the GPU unless its in-flight count exceeds the FPGA by tolerance $\delta$ and the FPGA has free capacity. If $\hat n_r<\theta$, it chooses the FPGA only when $q_{d_f^*}<c_{d_f^*}$ and FPGA load does not exceed GPU pressure; otherwise it falls back to the GPU. Explicit capacity checks prevent requests from accumulating behind a low-concurrency FPGA. Prefix precomputation is disabled for policies that create an FPGA queue, because KV can arrive while a request waits.

The decision is carried in request-level \texttt{kv\_transfer\_params}, including the target endpoint, a unique transfer ID, and prefix split. The prefill connector sends only to that endpoint and selects passthrough over an unrestricted cloud link for GPUs or RVQ over an independently rate-limited link for FPGAs. The transfer ID and prompt hash jointly match messages to requests under concurrency, duplicate prompts, and reordering.

\subsection{Execution-Level Optimizations}
\subsubsection{Prefill-Side Pipelined Transfer}
\begin{figure}[t]
 \centering
 \includegraphics[width=.9\columnwidth]{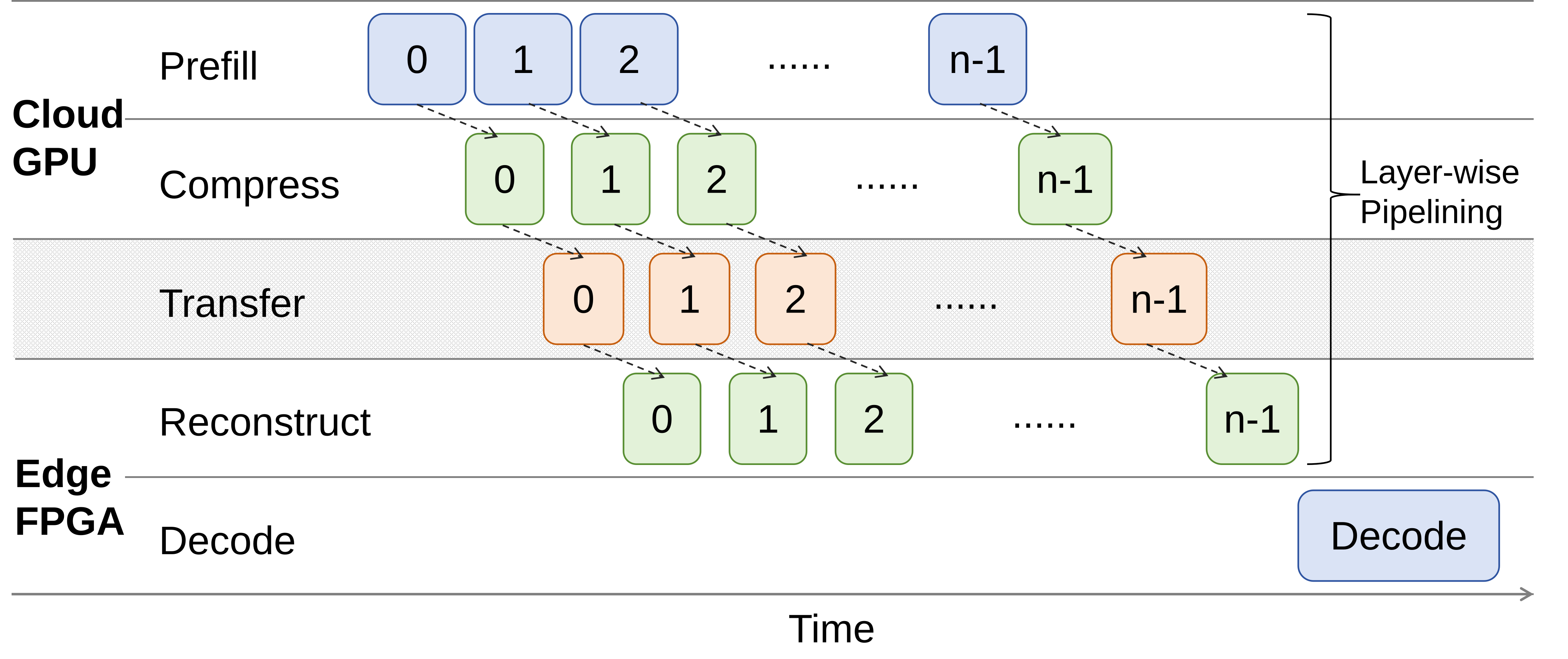}
 \caption{Layer-wise pipelining overlaps each KV layer's RVQ compression, device copy, and cloud--edge transfer with subsequent prefill layers.}
 \label{fig:pipeline}
\end{figure}

A naive PD path serializes $T^P+T^{enc}+T^{D2H}+T^{net}$, where D2H denotes a device-to-host copy. As Fig.~\ref{fig:pipeline} shows, since layer $l$ no longer modifies its KV after attention, \sys uses the vLLM \texttt{save\_kv\_layer()} callback to form a four-stage pipeline: extraction, RVQ encoding, D2H/bit packing, and transmission. After the main CUDA stream finishes layer $l$, the connector uses the PagedAttention slot mapping to extract request tokens and split keys and values. A dedicated encode stream waits for a production event, normalizes by standard deviation, and launches RVQ, while the main stream proceeds to layer $l+1$.

Background workers wait for encode events, copy only packed indices and scale factors into pinned CPU buffers, serialize them, and hand them to the transport layer. All GPU kernels are submitted by the vLLM worker's main thread; background threads handle only event waiting, D2H, serialization, and networking. Futures track outstanding layers. At prefill completion, \texttt{wait\_for\_save()} waits for all sends, emits a \texttt{done} message with the layer count, and creates a cache-ready marker, preventing a decoder from consuming a partial cache.

The connector groups heterogeneous destinations by codec and performs work only for the selected target. Lossless cloud endpoints complete synchronously, whereas edge sends may proceed asynchronously so that a slow wide-area link does not block the GPU path. Consequently, each layer's encoding, host processing, and transfer overlap with subsequent prefill computation. As Fig.~\ref{fig:pipeline} illustrates, the critical path is governed by the slower pipeline stage, apart from startup and tail costs.

\subsubsection{Decode-Side Precomputation}
\sys uses two complementary mechanisms. First, a receive thread starts before decode forward. Each arriving layer is deserialized, reconstructed, and written to the corresponding KV-cache region in memory. Per-layer events let the model call only \texttt{wait\_for\_layer\_load(l)} before layer $l$, overlapping later-layer reception and reconstruction with earlier computation. Messages arriving before local scheduling are retained under a prompt-hash and endpoint/transfer-ID key. An event fires only after every request in the current batch has written that layer's KV state; timeouts, unknown layers, and incomplete layer counts are detected explicitly.

\begin{figure}[t]
\centering
\includegraphics[width=.9\columnwidth]{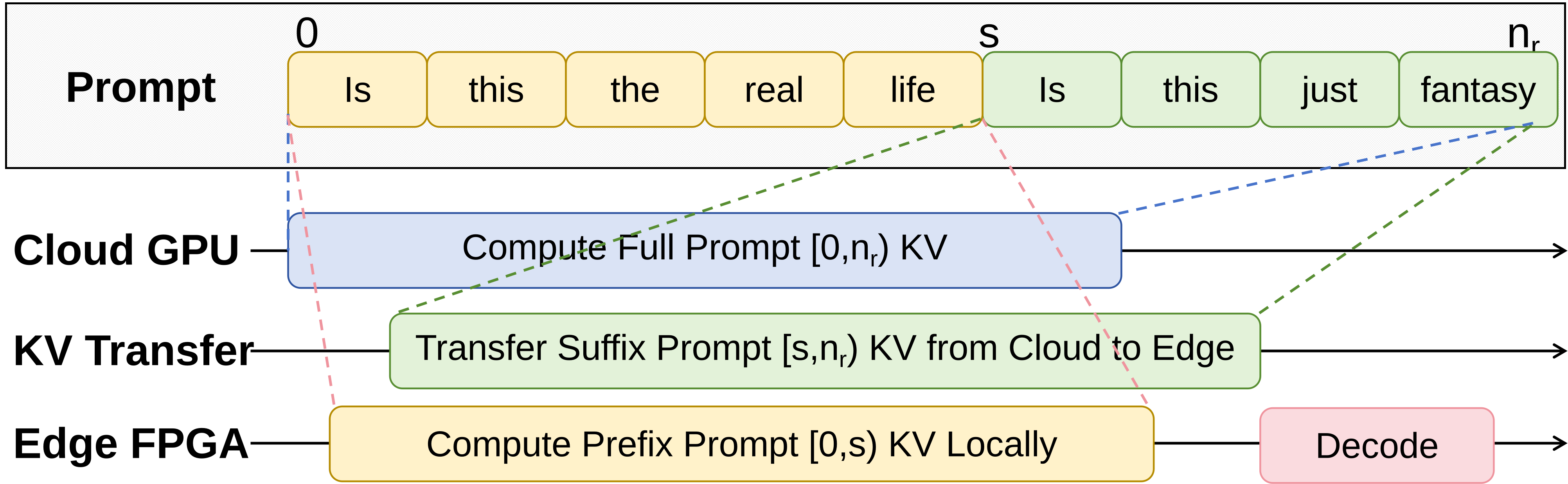}
\caption{Opportunistic decoder-side precomputation under low edge load. The FPGA computes a prompt prefix locally while receiving cloud-generated suffix KV.}
\label{fig:precompute}
\end{figure}

Second, edge-prefix precomputation (Fig.~\ref{fig:precompute}) targets intermittent edge arrivals, where the FPGA may frequently be idle. Under conventional PD disaggregation, when a request reaches an idle edge node, the FPGA must still wait for the cloud to generate and transfer its KV cache before decoding can begin. \sys uses this waiting window to compute a prompt prefix, overlapping otherwise idle FPGA work with cloud KV generation and transfer while reducing wide-area traffic. This design is inspired by Jin et al.'s concurrent KV computation and loading~\cite{jin2025computeloadkvcache}; \sys applies a similar compute--transfer principle across the cloud--edge boundary by overlapping local prefix computation with suffix-KV transfer.

For a request assigned to an idle FPGA, the GPU still prefills the complete prompt but transmits KV only for suffix $[s,n_r)$, while the FPGA computes the first $s$ tokens locally. The two paths start concurrently, and generation begins after both finish, so readiness is determined by the slower path. Here $0\le s<n_r$; GPU decoders always use $s=0$ and receive full lossless KV, whereas FPGA decoding receives only suffix KV beginning at token $s$.

Edge-prefix precomputation is an opportunistic mechanism for low-load operation. When the edge is idle, it uses the otherwise unoccupied compute window before KV arrival to generate prefix KV locally, trading local computation for fewer cross-domain bytes. Even for an idle FPGA, \sys enables this path only when the sequence is sufficiently long and bandwidth sufficiently low for the saved KV-transfer time to exceed the additional path overhead. Under sustained load, FPGA compute should instead prioritize requests already in decoding; \sys then sets $s=0$ and falls back to conventional PD-disaggregated execution.

\subsection{Bandwidth-Adaptive KV-Cache Compression}
\subsubsection{Residual Vector Quantization}
\begin{figure}[t]
\centering
\includegraphics[width=.78\columnwidth]{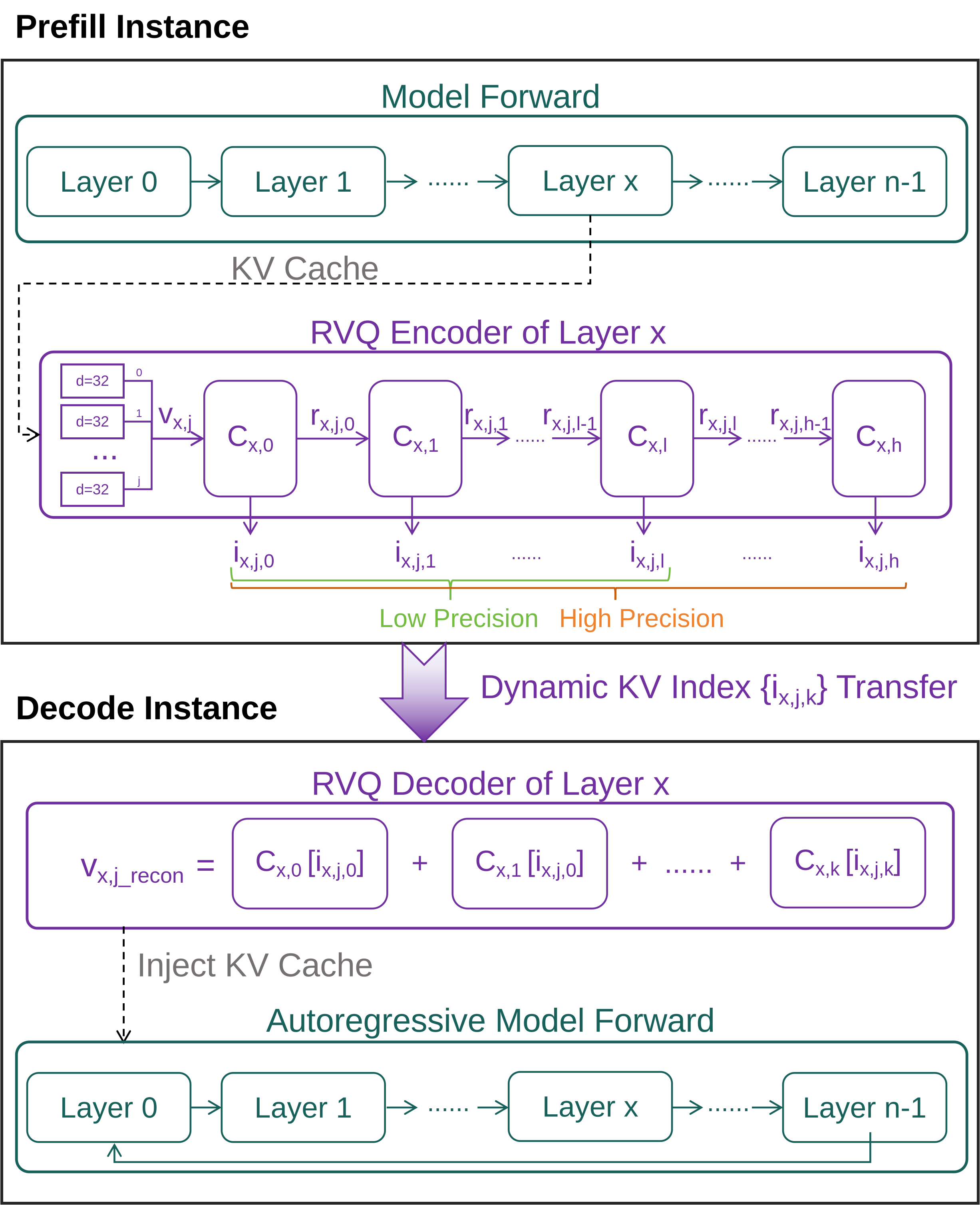}
\caption{RVQ-based KV-cache encoding at the cloud prefill worker and reconstruction at the edge decoder. Different residual depths provide low- and high-fidelity representations.}
\label{fig:rvq-design}
\end{figure}

Fig.~\ref{fig:rvq-design} illustrates cloud-side RVQ encoding and edge-side KV reconstruction. For a key or value vector $\mathbf{x}\in\mathbb{R}^d$ at layer $l$, \sys first computes $\sigma=\mathrm{std}(\mathbf{x})$ and normalizes $\bar{\mathbf{x}}=\mathbf{x}/\max(\sigma,\epsilon)$. It partitions the result into $G=d/\hat d$ subvectors. At RVQ stage $k$, each subvector $\mathbf z$ selects a layer-specific codeword:
\begin{equation}
\begin{split}
i_k&=\arg\min_{i\in[1,C]}\|\mathbf r_{k-1}-\mathbf c_{l,k,i}\|_2^2,\\
\mathbf r_k&=\mathbf r_{k-1}-\mathbf c_{l,k,i_k},\quad \mathbf r_0=\mathbf z,\\
\hat{\mathbf z}_K&=\sigma\sum_{k=1}^{K}\mathbf c_{l,k,i_k}.
\end{split}
\label{eq:rvq}
\end{equation}
The first $K$ stages of one $K_{\max}$-stage codebook therefore form a progressive representation: increasing $K$ reduces residual error but linearly increases index traffic.

\sys uses $C=1024$, $\hat d=32$, and $K_{\max}=16$, so each index occupies 10 bits. A Triton kernel packs four indices into five bytes on the GPU and transfers only packed indices and FP32 scale factors to the CPU. Ignoring small headers, one layer's key/value payload is
\begin{equation}
S(n,K)\approx 2n(G\cdot 10K/8+4).
\label{eq:rvq-size}
\end{equation}
Each key or value vector carries $G$ groups of indices and one 4-byte FP32 scale factor. This GPU-first path also avoids copying full FP16/BF16 KV tensors over PCIe.

Although VQLLM and related work have explored vector quantization for KV-cache compression~\cite{kumarResidualVectorQuantization2024}, they use it at a different system boundary from \sys. These approaches primarily target the KV memory footprint and retain the compressed representation throughout autoregressive generation, where frequent codec invocations introduce additional inference overhead. In contrast, \sys encodes and decodes each prompt cache once, only at the KV-migration boundary introduced by PD disaggregation, without changing ordinary KV accesses during subsequent decoding. Over a bandwidth-constrained cloud--edge link, this one-time codec cost is small relative to the transfer time saved by compression.

More importantly, RVQ's residual stages form a nested representation: the runtime can truncate one codebook stack after its first $K$ stages, incrementally adjusting traffic and reconstruction quality without changing the coding format or retraining the quantizer. Common KV-cache quantizers such as KIVI and KVQuant are designed around a precision selected in advance and do not natively treat compression depth as a control variable that follows link conditions. \sys therefore adopts RVQ not to claim uniformly higher static accuracy than every quantizer, but to obtain bandwidth adaptivity while retaining comparable model quality.

CacheGen also targets KV-cache network transfer, but relies on CUDA-based arithmetic coding for encoding and decoding~\cite{liuCacheGenKVCache2024}. Our measurements find its decoding very slow after migration to the edge platform, making it unsuitable for the cloud--edge KV path. RVQ reconstruction instead consists of regular codebook lookups and accumulation and provides substantially faster edge-side decoding.

\subsubsection{Bandwidth-Based Depth Selection}
Given bandwidth $B$, candidate depths $\mathcal K$, and quality floor $Q_{\min}$, the ideal choice is
\begin{equation}
\begin{split}
K^*(B,n)=\max\{K\in\mathcal K\mid\;&Q(K)\ge Q_{\min},\\
&T_{enc}(K)+S(n,K)/B\\
&\quad+T_{dec}(K)\le T_{budget}\}.
\end{split}
\label{eq:dstar}
\end{equation}
\sys builds a discrete operating table from offline quality tests and codec microbenchmarks and selects an entry according to the current link tier. Compressed messages carry shape, depth, scales, and indices; each FPGA endpoint has an independent serial send queue and bandwidth budget.

Compression depth and prefix split are jointly optimized. Under the quality constraint, \sys selects $(K,s)$ by balancing codec overhead, suffix-KV transfer time, and edge-prefix computation time, so that RVQ depth and prefix splitting share the same latency objective.

\section{Implementation}
The cloud implementation builds on vLLM's prefill--decode deployment and KV Connector. Requests assigned to GPU decoders reuse the existing intra-cloud KV-transfer path and high-speed interconnect, carrying lossless KV without an additional compression protocol. \sys extends only the edge-bound path: the connector invokes the RVQ kernel, issues asynchronous GPU-to-CPU copies, and hands compressed layers to the cloud--edge transport. A coordinator maintains performance profiles, organizes workers according to the FPGA-aware PD partition, and routes requests.

The edge service receives compressed layers, reconstructs their KV vectors from resident codebooks, and invokes the FlightLLM kernel deployed on the FPGA through OpenCL. The default two-GPU configuration uses two NVIDIA Tesla V100-PCIe GPUs with 32 GB memory and one AMD Xilinx Alveo U280: one V100 acts as the prefill worker and the other as the cloud decoder. We shape cloud--edge bandwidth to 100, 500, or 1000 Mbps. Table~\ref{tab:setup} summarizes the main setup.

Using performance data from the edge platform and network links, \sys dynamically adjusts request dispatch and transmitted KV precision during online serving to increase the likelihood of meeting SLOs. When it detects an extreme condition, such as a sudden bandwidth drop or an unresponsive FPGA, the service degrades gracefully and reschedules failed requests to a GPU for retry, preserving basic service stability.

The cloud--edge data path depends on neither NCCL, a specialized interconnect, nor an accelerator-specific memory layout. The sender packages RVQ indices, scale factors, and the necessary layer and tensor metadata into self-describing layer-wise messages. The receiver reconstructs KV from this representation and writes it into its own KV-cache memory.

\begin{table}[t]
\caption{System Configuration}
\label{tab:setup}
\centering
\small
\resizebox{\columnwidth}{!}{\begin{tabular}{ll}
\toprule
Component & Configuration \\
\midrule
Cloud accelerators & 2--4$\times$ Tesla V100-PCIe-32GB \\
Edge accelerator & AMD Xilinx Alveo U280 \\
Cloud runtime & vLLM with extended KV Connector \\
Edge runtime & FlightLLM \\
Link rates & 100, 500, 1000 Mbps \\
\bottomrule
\end{tabular}}
\end{table}

\section{Evaluation}

\subsection{Experimental Setup}
All experiments use Llama-3.1-8B-Instruct~\cite{grattafioriLlama3Herd2024}; FlightLLM uses the quantized model required by its kernel. The RVQ model is derived from a residual-quantized variational autoencoder (RQ-VAE) implementation~\cite{leeAutoregressiveImageGeneration2022a}, with optimized encode/decode kernels and runtime-selectable depth. Because \sys uses vLLM for GPU-side inference, we avoid conflating framework differences with system gains by selecting two primary baselines built on the same runtime: cloud-only vLLM using its example connector for disaggregated execution, and an unoptimized collaborative deployment combining native vLLM with FlightLLM. We additionally evaluate model-parallel Hugging Face Transformers~\cite{wolf-etal-2020-transformers} as a performance reference, wrapping it with a request queue and simple concurrent online service. All methods use the same GPUs and consistent optimization settings; collaborative methods additionally use the U280.

IMDB~\cite{maasLearningWordVectors2011}, CNN/DailyMail~\cite{hermannTeachingMachinesRead2015}, Qasper~\cite{dasigiDatasetInformationSeekingQuestions2021}, and GovReport~\cite{huangEfficientAttentionsLong2021} span short-text sentiment classification, medium summarization, long-document question answering, and long-form summarization. Table~\ref{tab:slo} summarizes their sequence lengths, request rates, concurrency, and TTFT SLOs; all use a 100-ms TPOT SLO. Request goodput is the number of completed requests satisfying both SLOs divided by measurement duration, with each request counted once regardless of token count. Other metrics are mean end-to-end latency, TTFT, per-request energy, and downstream accuracy. Transformers runs out of memory on GovReport and is omitted there.

\begin{table}[t]
\caption{Serving Workloads and Stress-Test Settings}
\label{tab:slo}
\centering
\scriptsize
\resizebox{\columnwidth}{!}{\begin{tabular}{lrrrrr}
\toprule
Dataset & Input & Output & Rate & Conc. & TTFT SLO \\
 & (tokens) & (tokens) & (qps) & & (ms) \\
\midrule
IMDB & 100--200 & 1--5 & 1 & 8 & 200 \\
CNN/DailyMail & $\sim$800 & $\sim$50 & 0.5 & 4 & 2,000 \\
Qasper & 5K--16K & 10--100 & 0.1 & 4 & 10,000 \\
GovReport & $<$9K & $<$500 & 0.05 & 4 & 20,000 \\
\bottomrule
\end{tabular}}
\end{table}

\begin{figure*}[t]
\centering
\includegraphics[width=.9\textwidth]{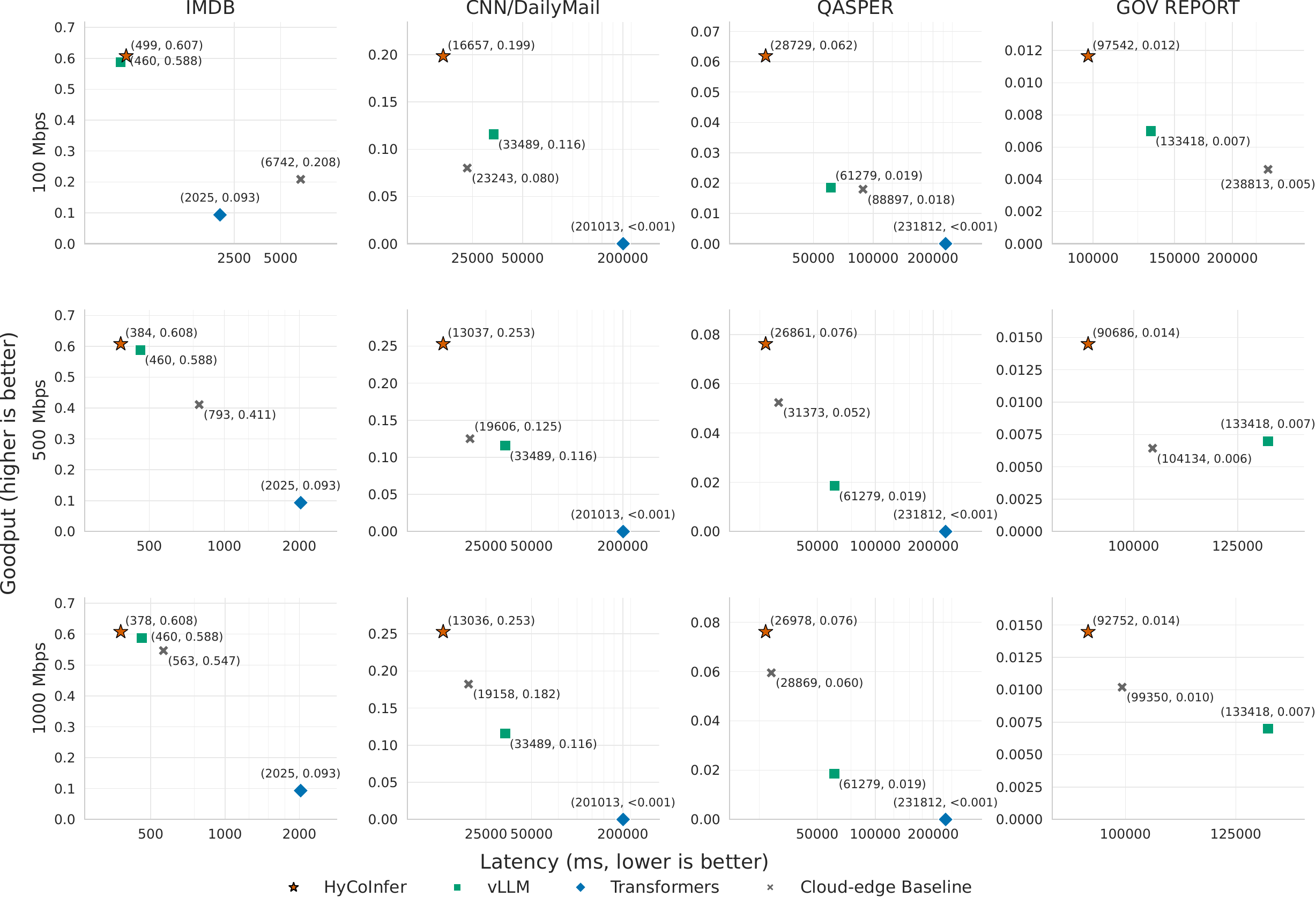}
\caption{SLO-compliant request goodput (requests/s) versus mean end-to-end latency for all workloads and cloud--edge bandwidths. Upper-left is better.}
\label{fig:endtoend}
\end{figure*}

\subsection{End-to-End Serving Performance}
Fig.~\ref{fig:endtoend} plots request goodput against mean latency; points toward the upper-left are better. \sys shows particularly strong gains on communication-intensive workloads. At 500 Mbps, it achieves 2.18$\times$, 4.11$\times$, and 2.07$\times$ the request goodput of cloud-only vLLM on CNN/DailyMail, Qasper, and GovReport, respectively, while reducing mean latency by 61.1\%, 56.2\%, and 32.0\%. Offloading relieves the GPU decode queue, allowing the cloud to accept new prefill requests sooner, while efficient FPGA decoding increases aggregate serving capacity.

The advantage over naive collaboration is especially clear under constrained bandwidth. At 100 Mbps, \sys achieves 3.45$\times$ the request goodput and 67.7\% lower latency on Qasper, and 2.52$\times$ the request goodput and 59.2\% lower latency on GovReport, relative to the unoptimized cloud--edge deployment. Naive offloading transfers the original KV cache and allows migration to dominate TTFT; \sys jointly controls this cost through compression, pipelining, and request dispatch.

From 500 to 1000 Mbps, \sys changes little while naive collaboration remains bandwidth-sensitive, indicating that compression and pipelining hide most transfer overhead. Selective offloading on Qasper and GovReport also frees GPU KV capacity and decode slots, sustaining gains without a high-bandwidth cloud--edge interconnect.

\begin{figure}[t]
\centering
\includegraphics[width=.95\columnwidth]{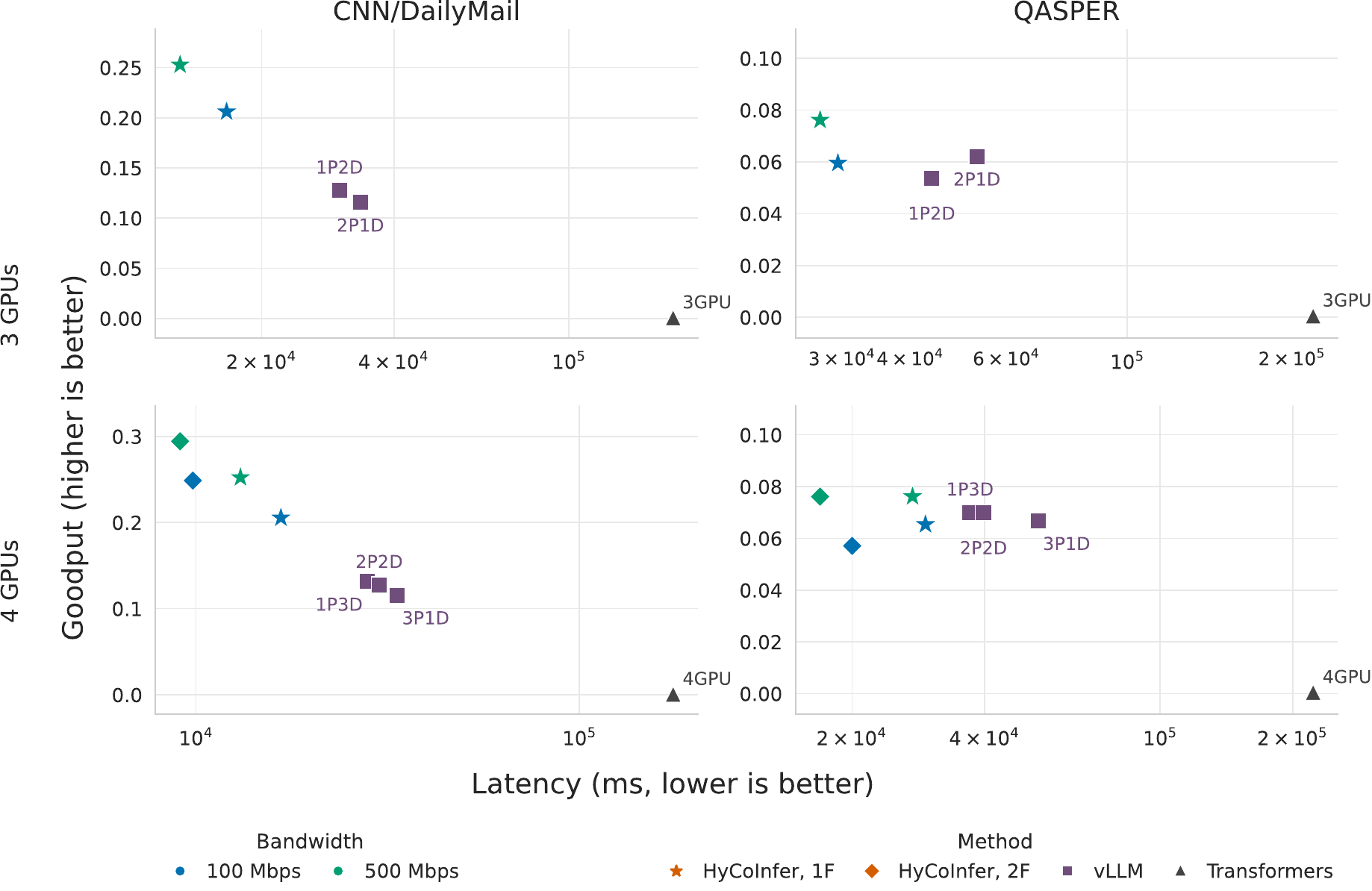}
\caption{Request goodput versus mean latency when scaling the cloud pool to three and four GPUs on CNN/DailyMail and Qasper. Upper-left is better.}
\label{fig:gpu-scale}
\end{figure}

\subsection{Multi-GPU Scalability}
Fig.~\ref{fig:gpu-scale} compares \sys with alternative cloud-only vLLM partitions and Transformers on CNN/DailyMail and Qasper as the cloud GPU pool expands. Across both workloads and resource scales, \sys maintains a favorable request-goodput/latency tradeoff. This result shows that FPGA-aware PD balancing scales with the heterogeneous resource pool by matching aggregate prefill and decode capacity, rather than relying on one fixed resource configuration.

\begin{figure}[t]
\centering
\includegraphics[width=.9\columnwidth]{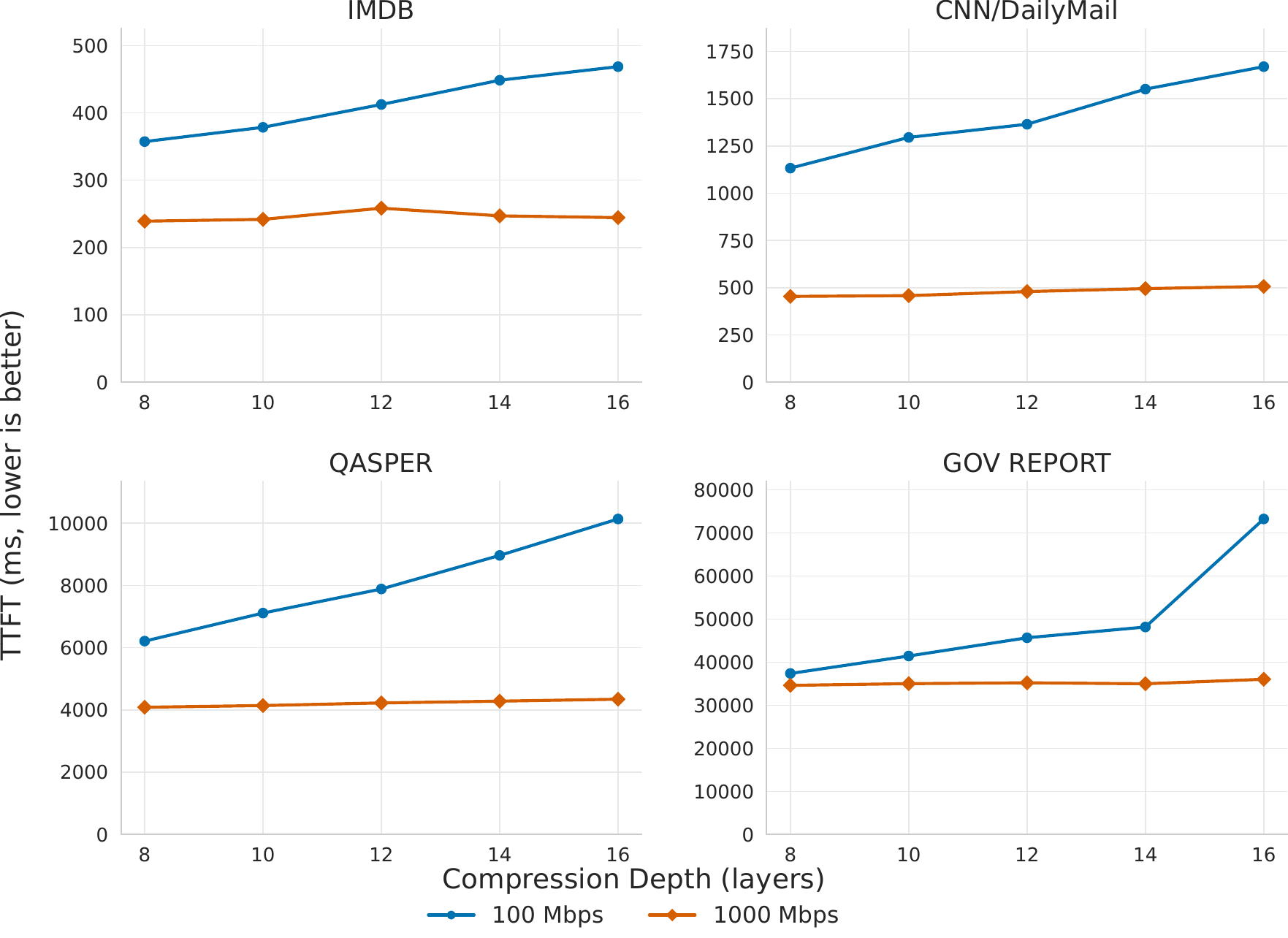}
\caption{TTFT versus RVQ depth under 100 and 1000 Mbps; lower is better.}
\label{fig:bw}
\end{figure}

\begin{figure}[!t]
\centering
\includegraphics[width=.9\columnwidth]{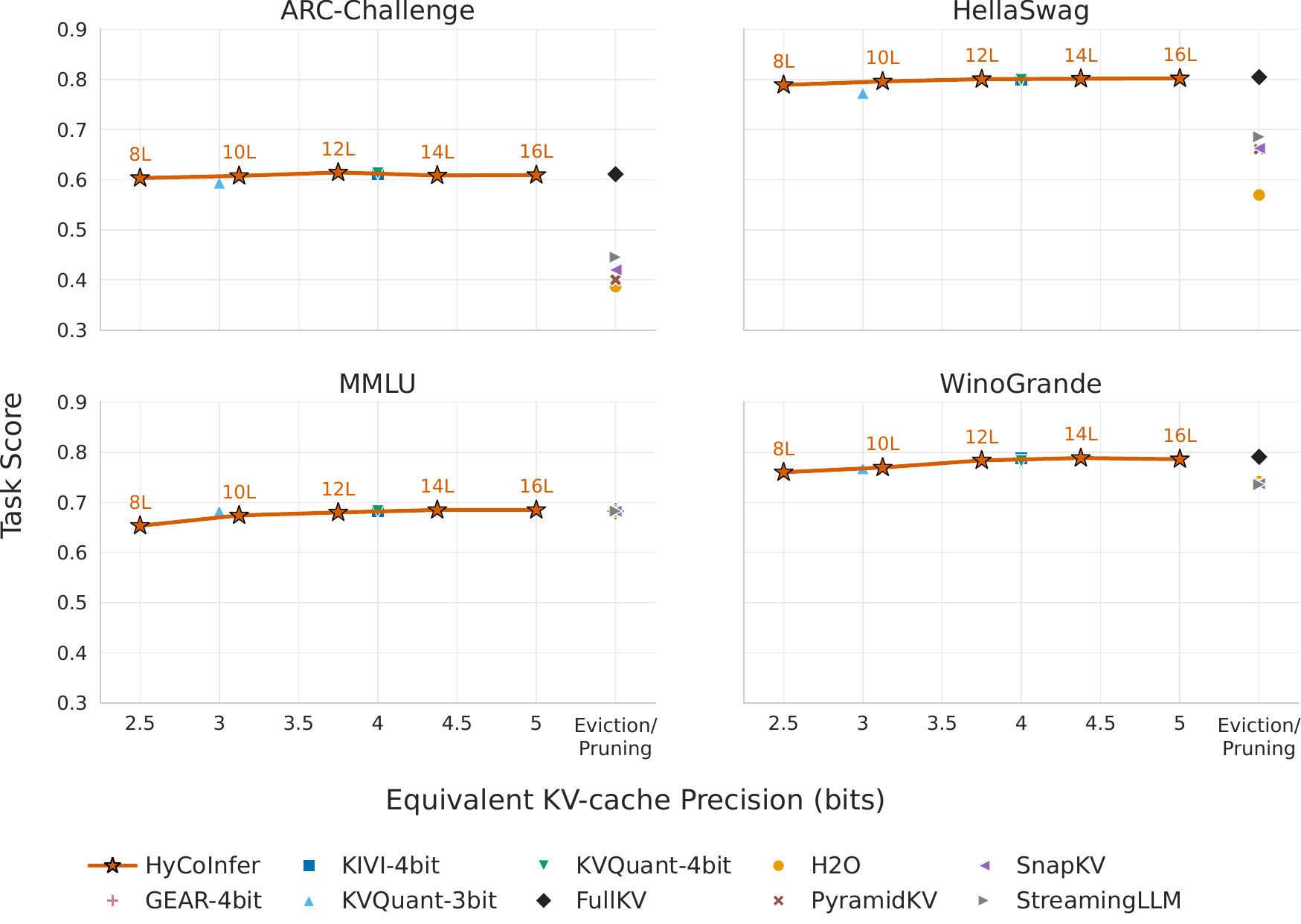}
\caption{Downstream quality (higher is better) of \sys at five RVQ depths, fixed-precision quantizers, eviction/pruning methods, and uncompressed FullKV.}
\label{fig:quality}
\end{figure}

\subsection{Bandwidth Sensitivity}
Fig.~\ref{fig:bw} shows TTFT versus RVQ depth for all four workloads at 100 and 1000 Mbps. At 100 Mbps, TTFT strongly depends on depth: reducing RVQ from 16 to eight stages lowers TTFT by 23.7--49.0\% across the workloads. GovReport, Qasper, and CNN/DailyMail improve by 49.0\%, 38.7\%, and 32.2\%, respectively, demonstrating that bandwidth-adaptive depth directly shortens the request critical path under a constrained link.

At 1000 Mbps, the TTFT difference between eight and 16 stages shrinks to 2.2--10.4\%. Once transfer shortens, layer-wise pipelining hides more of it behind prefill, allowing \sys to use a deeper representation for better reconstruction quality without a substantial TTFT penalty. This result demonstrates the feasibility of deeper compression configurations at higher bandwidth and suggests that the evaluated link rates leave flexible headroom for the increased KV-transfer pressure of larger models.

\subsection{RVQ Model Quality}
Fig.~\ref{fig:quality} compares \sys with two classes of KV-cache compression baselines on ARC-Challenge, HellaSwag, MMLU, and WinoGrande~\cite{clarkARC2018,zellersHellaSwag2019,hendrycksMeasuringMassiveMultitask2021,sakaguchiWinoGrande2020}. The quantization baselines are 4-bit KIVI~\cite{liuKIVITuningFreeAsymmetric2024}, 3/4-bit KVQuant~\cite{hooperKVQuant2024}, and 4-bit GEAR~\cite{kangGEAR2024}; the eviction or pruning baselines are H$_2$O~\cite{zhangH2OHeavyHitterOracle2023}, SnapKV~\cite{liSnapKVLLMKnows2024}, PyramidKV~\cite{caiPyramidKVDynamicKV2024}, and StreamingLLM~\cite{xiaoEfficientStreamingLanguage2024}. FullKV provides the uncompressed reference. The 8, 10, 12, 14, and 16 RVQ stages of \sys correspond to approximately 2.5, 3.125, 3.75, 4.375, and 5 index bits per KV element, excluding per-vector scale metadata.

RVQ depth correlates with task score, yet \sys retains high quality even at eight stages; when bandwidth permits, deeper representations move further toward FullKV. At similar equivalent precision, \sys provides overall task quality comparable to or slightly better than conventional quantizers, while eviction and pruning show more visible degradation on several tasks. These results show that RVQ supplies multiple useful compression operating points with limited quality loss. The key distinction is that a single codebook stack supports runtime truncation to different residual depths, allowing \sys to adapt traffic to bandwidth; fixed-precision methods such as KIVI and KVQuant do not natively provide this capability.

\subsection{Ablation Study}
Figs.~\ref{fig:ablation-pipeline}--\ref{fig:ablation-rvq} separately evaluate the three implemented optimizations across the four serving workloads.

\subsubsection{Layer-Wise Pipelining}
\begin{figure}[!t]
\centering
\includegraphics[width=\HyCoInferAblationWidth\columnwidth]{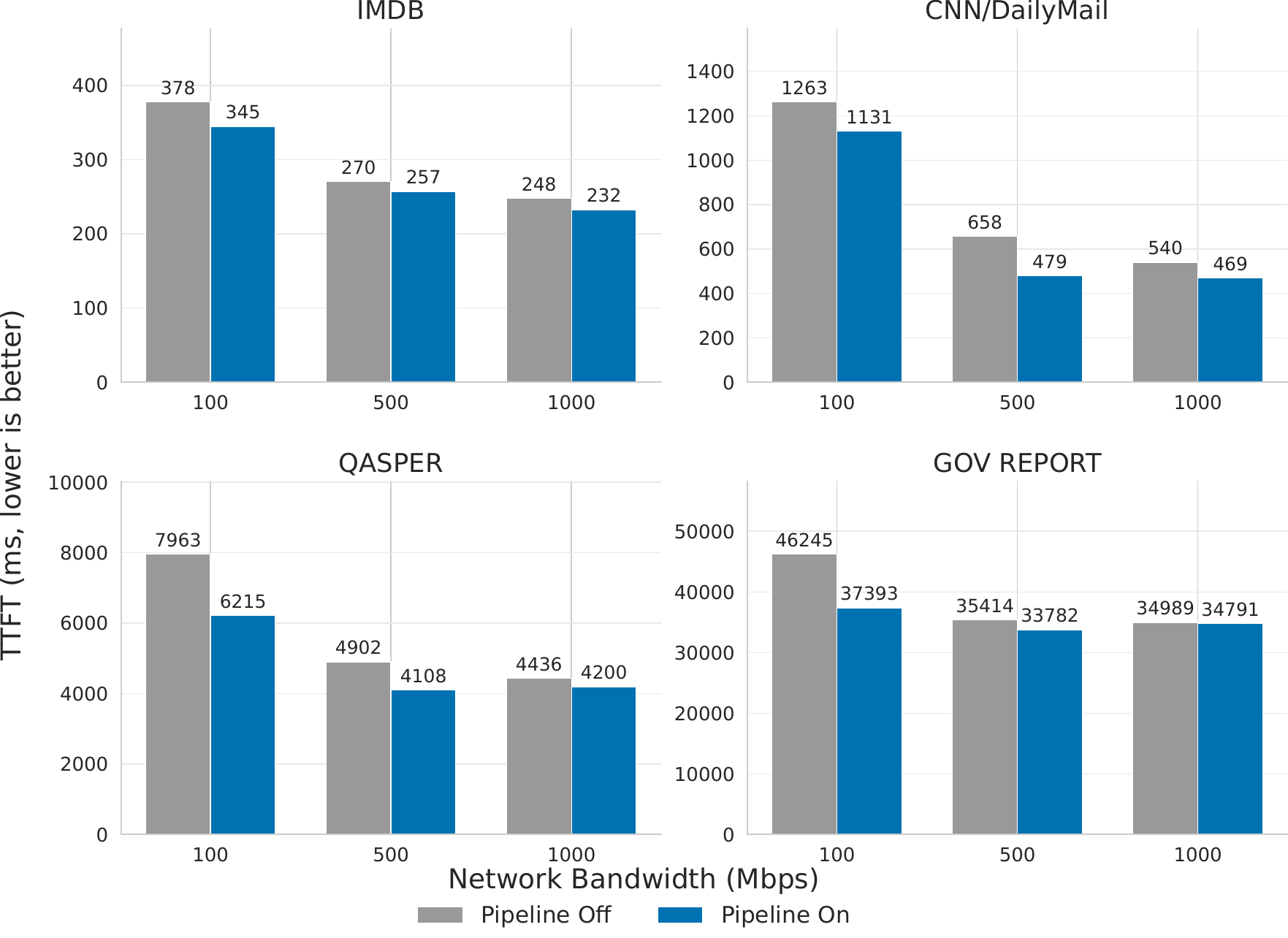}
\caption{TTFT ablation of layer-wise KV-cache pipelining across serving datasets and link rates; lower is better.}
\label{fig:ablation-pipeline}
\end{figure}

Fig.~\ref{fig:ablation-pipeline} shows that layer-wise pipelining lowers TTFT in every setting, by up to 27.2\%. At 100 Mbps, Qasper and GovReport improve by 22.0\% and 19.1\%, showing that later-layer computation hides earlier KV copy and transfer; CNN/DailyMail also improves by 27.2\% at 500 Mbps.

\subsubsection{Prompt-Aware Request Dispatch}
\begin{figure}[!t]
\centering
\includegraphics[width=\HyCoInferAblationWidth\columnwidth]{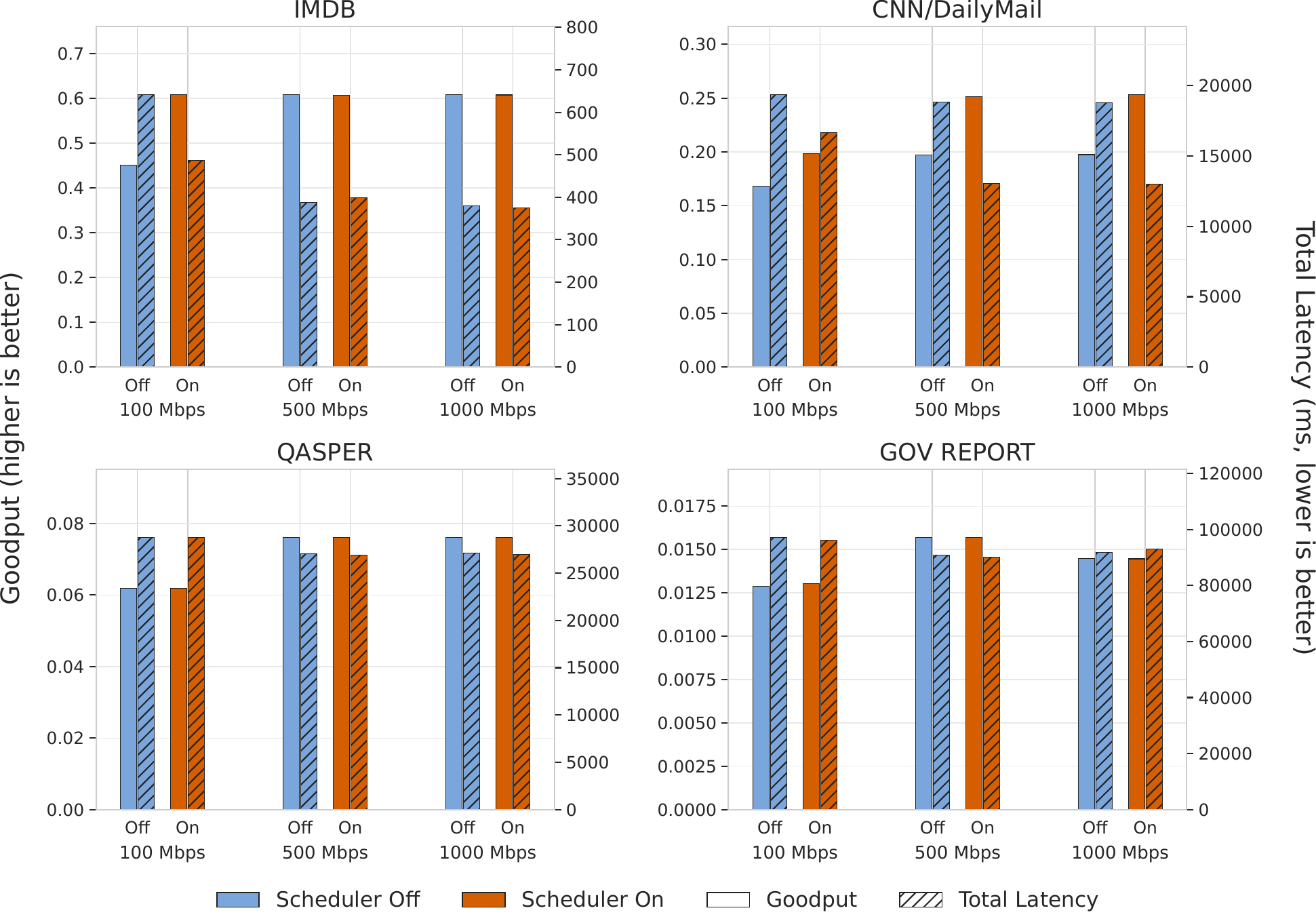}
\caption{Ablation of prompt-aware dispatch using SLO-compliant request goodput (requests/s) and mean latency; higher goodput and lower latency are better.}
\label{fig:ablation-dispatch}
\end{figure}

In Fig.~\ref{fig:ablation-dispatch}, dispatch improves CNN/DailyMail goodput by 18.0--28.2\% and lowers latency by 13.8--30.7\% across link rates; at 100 Mbps, it also improves IMDB goodput by 34.8\% and lowers latency by 24.0\%. It retains communication-expensive requests on GPUs while using idle FPGA capacity for suitable requests.

\subsubsection{RVQ Compression}
\begin{figure}[!t]
\centering
\includegraphics[width=\HyCoInferAblationWidth\columnwidth]{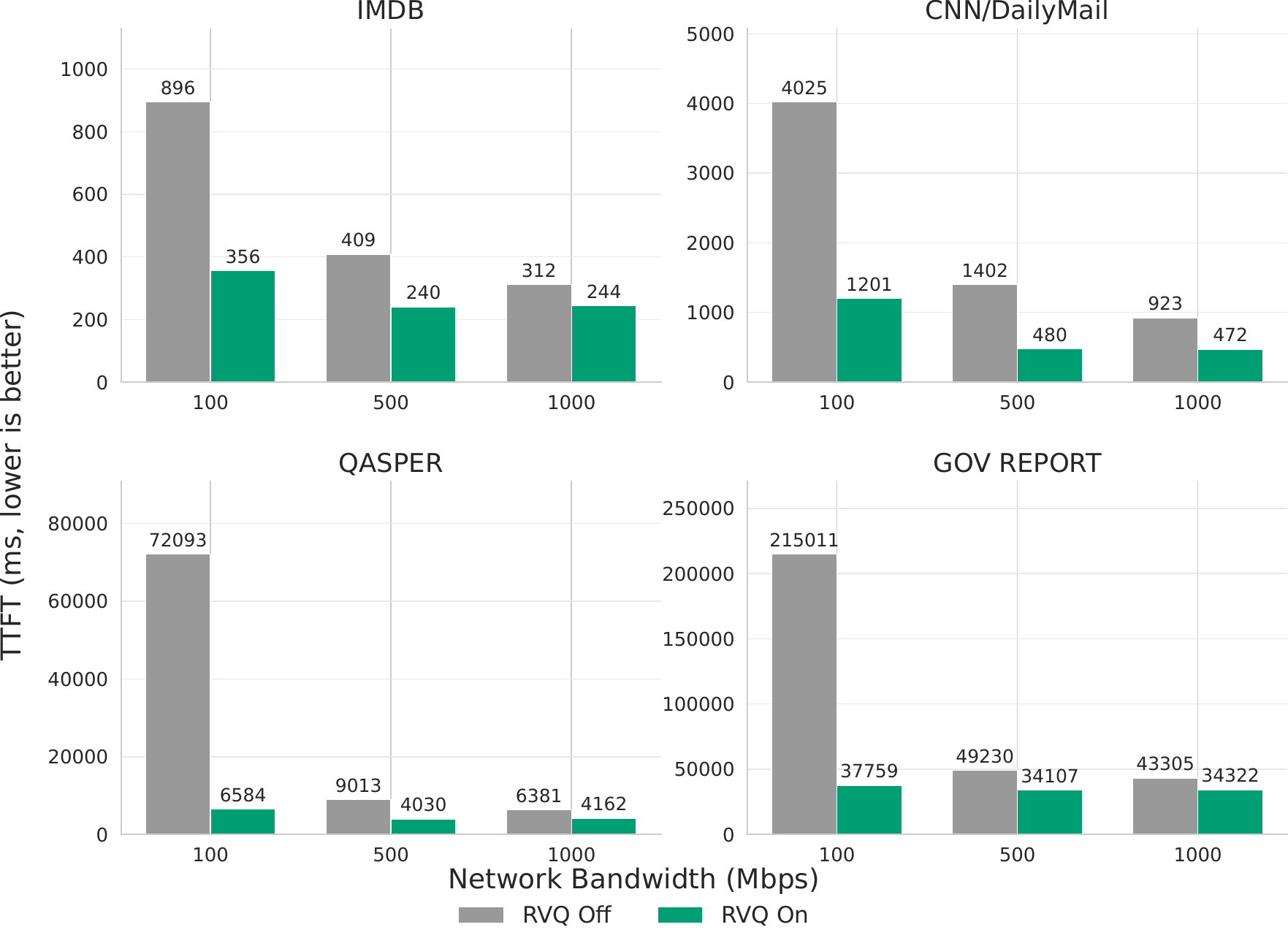}
\caption{TTFT ablation of RVQ-based KV-cache compression across serving datasets and link rates; lower is better.}
\label{fig:ablation-rvq}
\end{figure}

In Fig.~\ref{fig:ablation-rvq}, RVQ reduces TTFT by 60.2--90.9\% at 100 Mbps (90.9\% on Qasper and 82.4\% on GovReport), 30.7--65.8\% at 500 Mbps, and 20.7--48.9\% at 1000 Mbps, mitigating the low-bandwidth bottleneck while remaining useful as capacity rises.

\subsubsection{Edge-Prefix Precomputation}
\begin{figure}[!t]
\centering
\begin{minipage}[t]{.45\columnwidth}
\centering
\includegraphics[width=\linewidth]{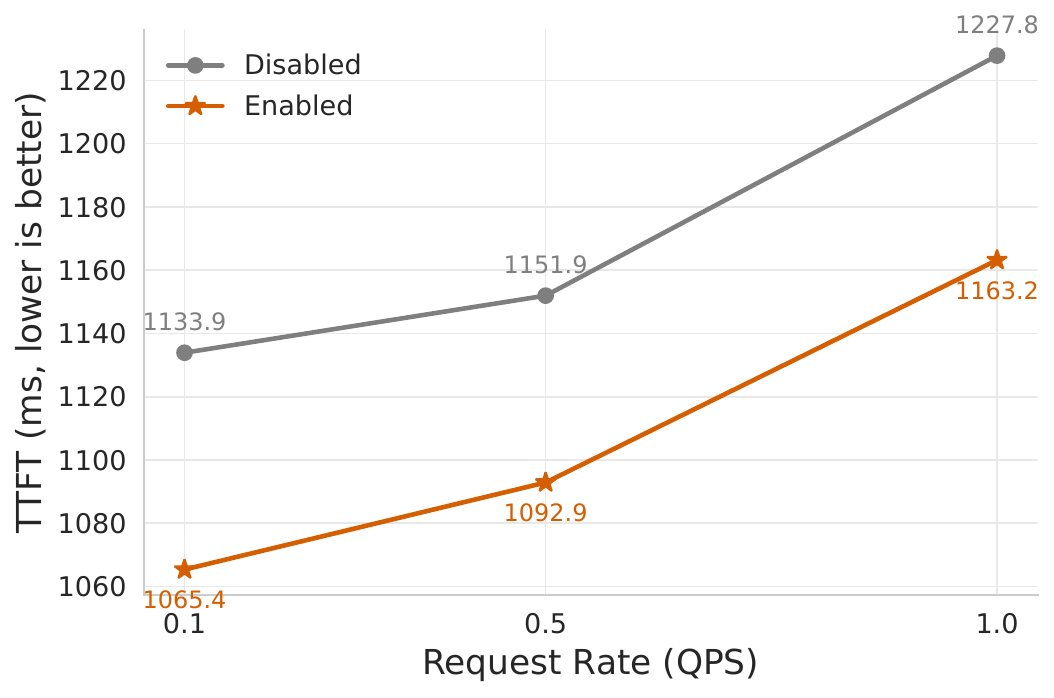}
\caption{TTFT with edge-prefix precomputation enabled or disabled across request rates; lower is better.}
\label{fig:precompute-ablation}
\end{minipage}\hfill
\begin{minipage}[t]{.45\columnwidth}
\centering
\includegraphics[width=\linewidth]{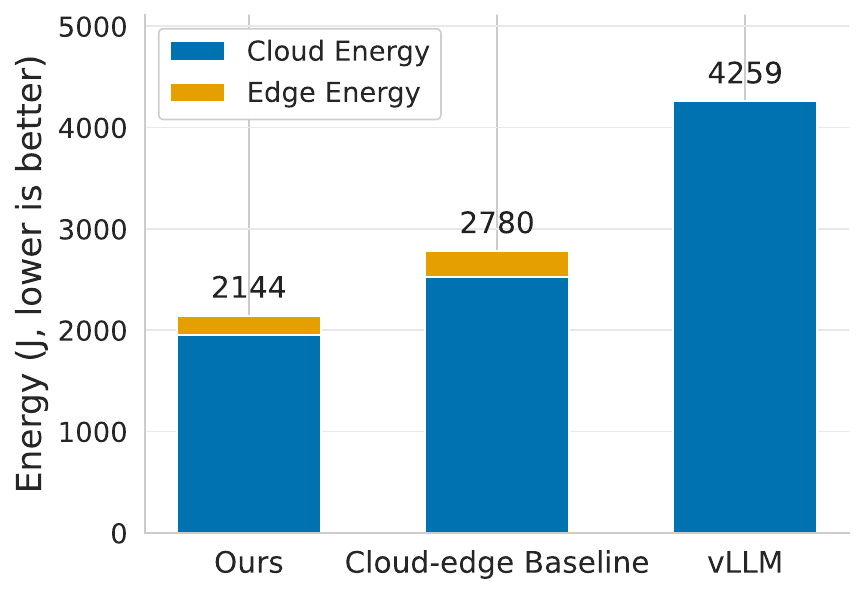}
\caption{Estimated cloud and edge accelerator energy per request; lower is better.}
\label{fig:energy}
\end{minipage}
\end{figure}

In Fig.~\ref{fig:precompute-ablation}, each point averages six paired CNN/DailyMail runs. Across 0.1--1.0 QPS, precomputation reduces mean TTFT by 5.1--6.0\%, showing that idle edge compute can shorten KV readiness; under sustained load, \sys falls back to conventional PD disaggregation (Sec.~III-C2).

\subsection{Energy Efficiency}
Combining sampled accelerator power with completed-request rate, Fig.~\ref{fig:energy} shows that collaborative execution lowers estimated CNN/DailyMail energy per request over cloud-only serving. FPGA decode thus improves accelerator efficiency while releasing cloud capacity.

\FloatBarrier
\section{Related Work}
\subsection{Efficient LLM Serving}
vLLM~\cite{kwonEfficientMemoryManagement2023} and SGLang~\cite{zhengSGLangEfficientExecution2023} improve GPU serving through batching, memory management, and scheduling, while GPTQ~\cite{frantarGPTQAccuratePostTraining2022} and AWQ~\cite{linAWQActivationawareWeight2023} reduce model-weight traffic. These techniques are orthogonal to \sys: quantized models can run in either decoder pool, while our system decides where a request executes and how its KV state crosses the network.

\subsection{Prefill--Decode Disaggregation}
DistServe~\cite{zhongDistServeDisaggregatingPrefill2024} and Splitwise~\cite{patelSplitwiseEfficientGenerative2024} separate the two phases in GPU clusters to isolate interference and provision them independently; Mooncake, MemServe, and LMCache further study KV scheduling, reuse, and transfer within a cluster~\cite{qinMooncakeKVCachecentricDisaggregated2025d,huMemServeContextCaching2024,liuLMCacheEfficientKV2025}. \sys addresses a different regime: an FPGA decoder behind a constrained, variable edge link, where the first cross-domain migration of a new KV cache requires progressive compression and selective offloading.

\subsection{FPGA Inference}
FlightLLM demonstrates efficient LLM inference with customized FPGA dataflow and quantization~\cite{zengFlightLLMEfficientLarge2024a}; recent work also explores GPU--FPGA execution and scalable dataflows~\cite{yangGLITCHESGPUFPGALLM2024a,zhengLoopLynxScalableDataflow2025a}. \sys focuses on online integration: cloud GPUs perform prefill, FPGAs join the decoder pool, and routing jointly accounts for edge execution time, independent queues, and cross-domain transfer rather than assuming every offload is beneficial.

\subsection{KV-Cache Compression}
KV compression methods fall into several categories: H$_2$O and SnapKV selectively retain tokens~\cite{zhangH2OHeavyHitterOracle2023,liSnapKVLLMKnows2024}, while KIVI, KVQuant, GEAR, and ZipCache reduce numerical width~\cite{liuKIVITuningFreeAsymmetric2024,hooperKVQuant2024,kangGEAR2024,heZipCacheAccurateEfficient2024}. These methods generally use a fixed representation or quality policy. CacheGen supports dynamic compression through arithmetic coding~\cite{liuCacheGenKVCache2024}, but its edge decoding was too slow in our measurements. In contrast, \sys uses fast progressive RVQ, whose nested depth adapts to bandwidth without receiver changes, retraining, or token eviction.

\section{Conclusion}
\sys jointly optimizes scheduling and KV migration across GPUs and FPGAs. On V100-U280 at 500 Mbps, it delivers up to 4.11$\times$ the request goodput and 61.1\% lower mean latency than cloud-only vLLM while reducing energy per request.

\clearpage
\bibliographystyle{IEEEtran}
\bibliography{refs-infocom}

\end{document}